\documentclass[11pt]{article}
\usepackage{blois,epsfig,amsmath,amssymb}

\def\eg{{\it e.g.}}
\def\ie{{\it i.e.}}
\def\vev#1{\langle #1 \rangle}
\def\LTC{\Lambda_{\text{TC}}}
\begin{document}
\vspace*{4cm}
\title{STRONG ELECTROWEAK SYMMETRY BREAKING}

\author{BENJAMIN GRINSTEIN}

\address{UCSD, Physics Department, La Jolla, CA, 92093-0319, USA\\
and\\
CERN, Physics-Theory Group,  CH-1211, Geneva 23, Switzerland
}

\maketitle\abstracts{ Models of spontaneous breaking of electroweak
  symmetry by a strong interaction do not have fine tuning/hierarchy
  problem. They are conceptually elegant and use the only mechanism of
  spontaneous breaking of a gauge symmetry that is known to occur in
  nature. The simplest model, minimal technicolor with extended
  technicolor interactions, is appealing because one can calculate by
  scaling up from QCD. But it is ruled out on many counts:
  inappropriately low quark and lepton masses (or excessive FCNC), bad
  electroweak data fits, light scalar and vector states, etc. However,
  nature may not choose the minimal model and then we are stuck:
  except possibly through lattice simulations, we are unable to
  compute and test the models. In the LHC era it therefore makes sense
  to abandon specific models (of strong EW breaking) and concentrate
  on generic features that may indicate discovery. The Technicolor
  Straw Man is not a model but a parametrized search strategy
  inspired by a remarkable generic feature of walking technicolor,
  that technivector mesons are light, narrow and decay readily into
  electroweak vector mesons and photons. While walking technicolor is
  popular among practitioners, alternatives exist and the Straw Man may
  not lead to their discovery.}

\section{Introduction: Why Strong Electroweak Symmetry Breaking?}
Theories of strong electroweak symmetry breaking (SESB) do not include
fundamental scalar fields.  Masses of vector and spinor fields are
protected from divergent corrections automatically, those of vectors
by gauge invariance and those of spinors by chiral symmetry. 
Hence, at least in principle, such theories  elegantly and thriftily
solve the fine tuning problem. 

Moreover, they easily and naturally generate large hierarchies of
scales. These theories are asymptotically free. If $\Lambda_s$ denotes
the scale at which the running coupling $\alpha(\mu)$ becomes large,
inducing an electroweak symmetry breaking condensate, then roughly
\begin{equation}
  \Lambda_s=M\exp\left(\frac{8\pi/b}{\alpha(M)}\right)\,,
\end{equation}
where $b$ is the one loop coefficient of the beta function and $M$ is%
, say, the Planck mass. 

We don't know ``why'' nature would use one mechanism or another to
break a symmetry spontaneously. But we do know it did choose
non-perturbative condensates to break flavor symmetry, $SU(3)\times
SU(3)\to SU(3)$, and, according to the BCS theory, electron
condensates  to break $U(1)_{\text{em}}$. 

\paragraph{Disclaimer} We can classify SESB according to whether or
not the strong interaction produces a composite light scalar (\eg, a pseudo
goldstone boson) that acts as a higgs field. I had hoped to talk about
the case where there are higgs-like light bound states, but to avoid
duplication I agreed to talk about something I know less about,  higgsless models, while
Christophe Grogean's talk is on pseudo goldstone higgs particles. The
two talks roughly parallel the entries on SESB in the 2009 Les Houches LHC
report.\cite{Brooijmans:2010tn}

\subsection{Why not}
In addition to the inherent inability to calculate with them, a number
of shortcomings have made theories of SESB unpopular. The simplest
models of SESB, which we review below, have difficulty generating quark
and lepton masses, run into trouble with constraints from FCNCs and
electroweak precision data,  have very light spin-0 particles and it is
not apparent how, if at all,  can be incorporated into a GUT.

Several frameworks, like extended-, walking-, conformal- and top-color
assisted- technicolor, have been formulated to address these
problems. The plan of the talk is as follows: first review the minimal
technicolor model so we may better appreciate the ``why not's'' of
SESB. Then review some of the frameworks that address these. In
particular because of time and space limitations we shall focus on
extended- and walking- technicolor (find more details in lectures by
Ken Lane\cite{Lane:2002wv}). After evaluating these we will
proceed to briefly discuss LHC discovery prospects.

\section{Minimal Technicolor: A beautiful idea that does not work}
Two flavor QCD is a theory that well approximates the world at low
energies. It has an approximate chiral $SU(2)_L\times SU(2)_R$ symmetry
that is broken spontaneously to the vector (diagonal sum)  $SU(2)_V$
by a strong interaction condensate. In a bit more detail 
$q_{L(R)}=(u_{L(R)},d_{L(R)})^T$  is an $SU(2)_{L(R)}$ doublet, and
the strong interaction induces condensation:
\begin{equation}
\label{eq:QCDcond}
\vev{\bar q_L^i q_R^j} =\Lambda_\chi^3\delta^{ij}
\end{equation}
We interpret the resulting pseudo-Goldstone bosons as pions ($\pi^\pm,
\pi^0$), so the scale $\Lambda_\chi$ of the condensate must be related
to their decay constant, $f_\pi$. One typically estimates
$\Lambda_\chi\sim4\pi f_\pi$, or, perhaps, $\Lambda_\chi^3\sim4\pi
f_\pi^3$. Pions are not exact goldstone bosons. The quark masses
$m_{u,d}\ne0$ and electromagnetic couplings break the chiral symmetry
explicitly resulting in non-vanishing  pion masses. 

Recall that in the standard model (SM) of electroweak interactions an
$SU(2)_L\times U(1)_R$ subgroup of the chiral $SU(2)_L\times SU(2)_R$
is gauged.\footnote{Consistent quantization of the gauge symmetry
  requires that the symmetry be exact. Hence the explicit masses of
  quarks, $m_{u,d}$, have to be put to zero.} Even in the absence of a
higgs field, the symmetry of the SM is spontaneously broken by the QCD
condensate in \eqref{eq:QCDcond}. The would-be goldstone bosons are
``eaten,'' that is, they become the longitudinal components of the $W$
and $Z$.  The $W$ and $Z$ masses obtained this way are too small, of
the order of the pion decay constant. However, there is one successful
relation. In $SU(2)_L\times SU(2)_R\to SU(2)_V$ the group $SU(2)_V$ is
explicitly broken by the gauging of $U(1)_R$. Hence, up to hypercharge
corrections, this $SU(2)_V$ ``custodial symmetry'' guarantees
\begin{equation}
M_W=M_Z\cos\theta_W\,,\qquad\text{that is,}\qquad
\rho=\frac{M_W^2}{\cos^2\theta_W M_Z^2}=1\,.
\end{equation}

Minimal Technicolor\,\cite{TC} is a copy of 2-flavor QCD with:
\begin{itemize}
\item ``QCD'' replaced by ``TC,'' ``gluons'' by ``technigluons,''  ``quarks'' by
  ``techniquarks,'' etc.
\item The technistrong coupling constant taken so that the condensate is at the
  electroweak scale: $\LTC\sim4\pi
  v=4\pi(246~\text{GeV})$.
\item An $SU(2)_L\times U(1)_R$ subgroup of $SU(2)_L\times SU(2)_R$ 
  gauged giving electroweak interactions.
\end{itemize}
While we cannot calculate much from first principles in this
model, we can infer  a lot from experiment: since it is a scaled up copy of QCD
it suffices to measure QCD parameters and scale them by the
appropriate factor. This is what was done, for example, in order to
confront with electroweak data and conclude that minimal technicolor
is all but ruled out.\cite{Peskin:1990zt}

\section{Masses, Goldstone Bosons, Extended Technicolor and the Problem with FCNC}
The higgs field in the SM plays a dual role. On the one hand it breaks
electroweak symmetry, giving the $W$ and $Z$ masses, and on the other
its Yukawa interactions with quarks and leptons, $Y_U\tilde H\bar q_L
u_R+Y_D H\bar q_L d_R+Y_E H\bar \ell_L e_R$, produce their masses. In
the minimal TC model techniquark condensates $\vev{\bar Q_L Q_R}$
break electroweak symmetry and give masses to the $W$ and $Z$, but
quarks and leptons remain massless. Couplings of techniquarks to
quarks and leptons are needed to generate their masses.

Whatever new interaction is responsible for quark and lepton masses,
they may solve another problem with TC models. In the minimal TC model
above there are exactly three would-be goldstone bosons that are eaten
by the $W^\pm$ and $Z^0$. It is exactly three because the pattern of
symmetry breaking involved small groups, $SU(2)_L\times SU(2)_R\to
SU(2)_V$. Many different patterns of symmetry breaking are possible
and generically lead to more goldstone bosons.  Three are still eaten
by the $W/Z$ but the rest, the ``technipions,'' remain in the
spectrum.  Their masses arise from electroweak interactions. Hence
they are light and ruled out experimentally. The additional
interactions that give rise to quark masses may perhaps render the
technipions safely heavier.

In Extended TC (ETC)\,\cite{ETC} we assume there are additional
interaction terms of the form\,\footnote{There are alternatives to ETC
  for mass generation, like
  partial compositeness.\cite{Kaplan:1991dc} See the discussion of
  Minimal Composite Higgs\,\cite{Agashe:2004rs} in Grojean's talk, this volume.}
\begin{equation}
\label{eq:ETC}
\frac{C}{M^2}\bar Q Q \bar q_L q_R\qquad \text{and}\qquad \frac{C}{M^2}\bar Q Q \bar \ell_L e_R\,.
\end{equation}
These are operators of dimension 6 so they appear in the Lagrangian
with two inverse powers of a mass-dimensional parameter, here denoted
by $M$. The $C$ are dimensionless constants that characterize the quark mass
matrices (playing a role analogous to that of the $Y_Q$ of the
SM). We read off the quark and lepton masses:
\begin{equation}
\label{eq:qmass}
m_{q,\ell}\sim\frac{C}{M^2}\vev{\bar QQ}\sim\frac{C}{M^2}\LTC^3\,.
\end{equation}

\begin{figure}[t]
\begin{center}\includegraphics[height=4cm]{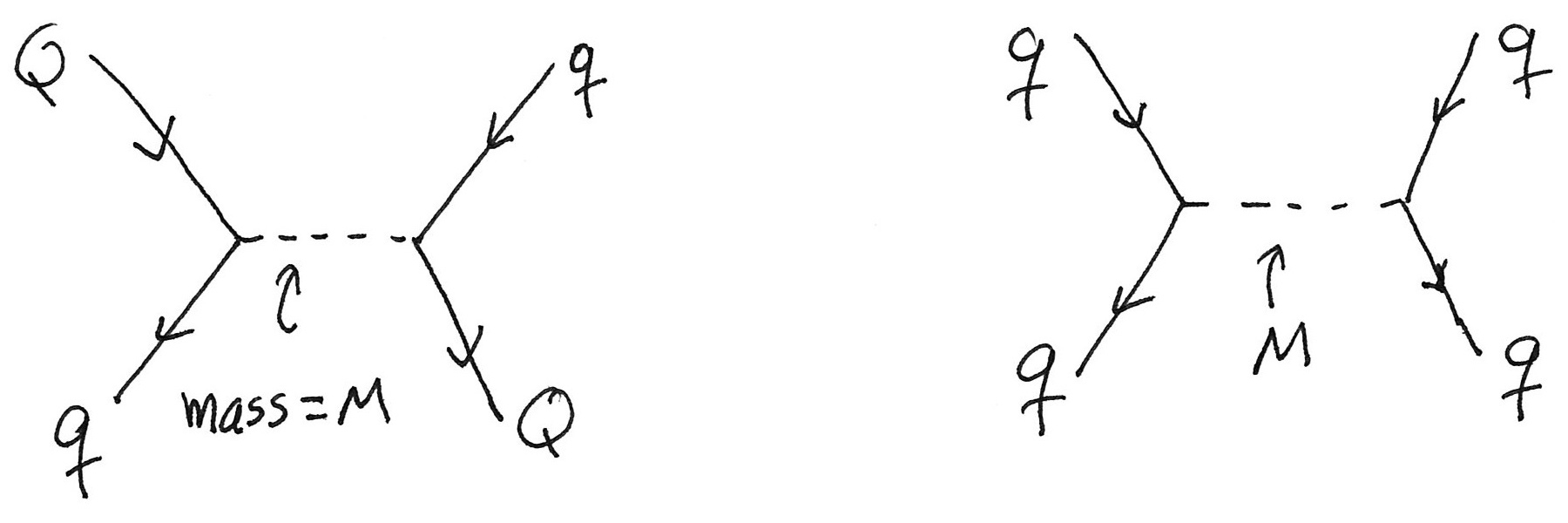}
\end{center}
\caption{Feynman diagrams of some Extended Technicolor
  interactions. At energies well below $M$ they give rise to
  four-fermion operators. The diagram on the left produces effective
  interactions like those in \eqref{eq:ETC}, responsible for quark and
  lepton masses. The diagram on the right induces troubling FCNCs.
\label{fig:etc-int}}
\end{figure}

The operators in \eqref{eq:ETC} are nothing but the effective
description of an interaction that arises from the exchange of a heavy
particle of mass $\sim M$. A complete ETC model should specify the new
interaction, but for our purposes it suffices to assume its
existence. The first of the Feynman diagrams in Fig.~\ref{fig:etc-int}
is an example of the exchange of a particle of mass $M$ that gives
rise to the effective interaction in \eqref{eq:ETC}. The second
Feynman diagram in the figure  shows  exchange of the
same heavy particle between quarks only,  using the same basic
vertices that were introduced in order to produce quark masses. The
inescapable conclusion is that  ETC interaction will generically 
produce four quark operators too, like 
\begin{equation}
\frac{C'}{M^2}\bar qq \bar q q\,,\qquad\text{and in particular $\Delta
  S=1$, FCNC terms like ~~}
\frac{C'}{M^2}\bar sd \bar sd\,.
\end{equation}
A quick and dirty bound on $C'/M^2$ can be found by requiring it does
not exceed  the corresponding coefficient of the four quark operator
in the SM.\footnote{This ignores actual top quark mass dependence and
  differences in the Dirac structure of the four quark operators, but
  it is good enough for order-of-magnitude estimates} Now, in the SM
this arises largely from a1-loop diagram, a  $W$-box with internal top
quarks. So we must have
\begin{equation}
\label{eq:fcnc}
\frac{C'}{M^2}\lesssim \frac{1}{16\pi^2}G_F |V_{td}V_{ts}|^2
\qquad\Rightarrow\qquad M/\sqrt{|C'|}\gtrsim 10^4~\text{TeV}
\end{equation}

Herein lies the problem: using the bound \eqref{eq:fcnc} in
\eqref{eq:qmass} and assuming roughly that $C'\approx C$ the largest possible
quark and lepton masses  are in the MeV range. One can try to
get around this problem by making $C'$ very small compared to
$C$, which requires creative model building. In particular, for a  technipurist
there ought not to exist any scalar fields:  the particle exchanged in
the ETC must be a heavy vector boson, presumably associated with some
non-abelian symmetry (that does not commute with neither  EW nor TC
interactions). Then the relative size of $C'$ and $C$ is fixed by
group theory and we expect $C'\sim C$.\footnote{An exception is if
  quarks and techniquarks belong in distinct representations of the
  ETC gauge group that are very different in size. But then it is
  difficult to maintain asymptotic freedom}

The ETC will also give  four-techniquark operators, $(C''/M^2)\bar Q
Q\bar QQ$. These generically break the global symmetries of the TC
theory and therefore contribute to the mass of technipions. But the
extremely high mass of the ETC scale renders the effect insufficient: roughly
$f^2_\pi m_\pi^2\sim C'' \LTC^6/M^2$, or $m_\pi\sim
4\pi\LTC^2/M\lesssim 10$~GeV.

\begin{figure}
\includegraphics[height=5cm]{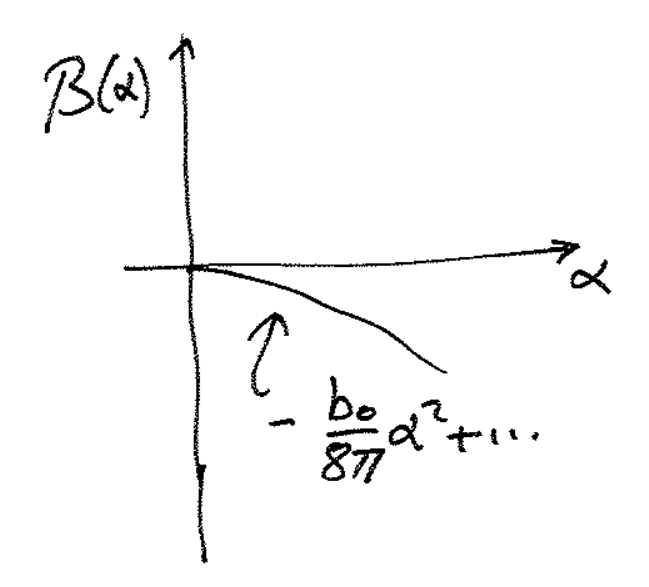}\hspace{3cm}
\includegraphics[height=5cm]{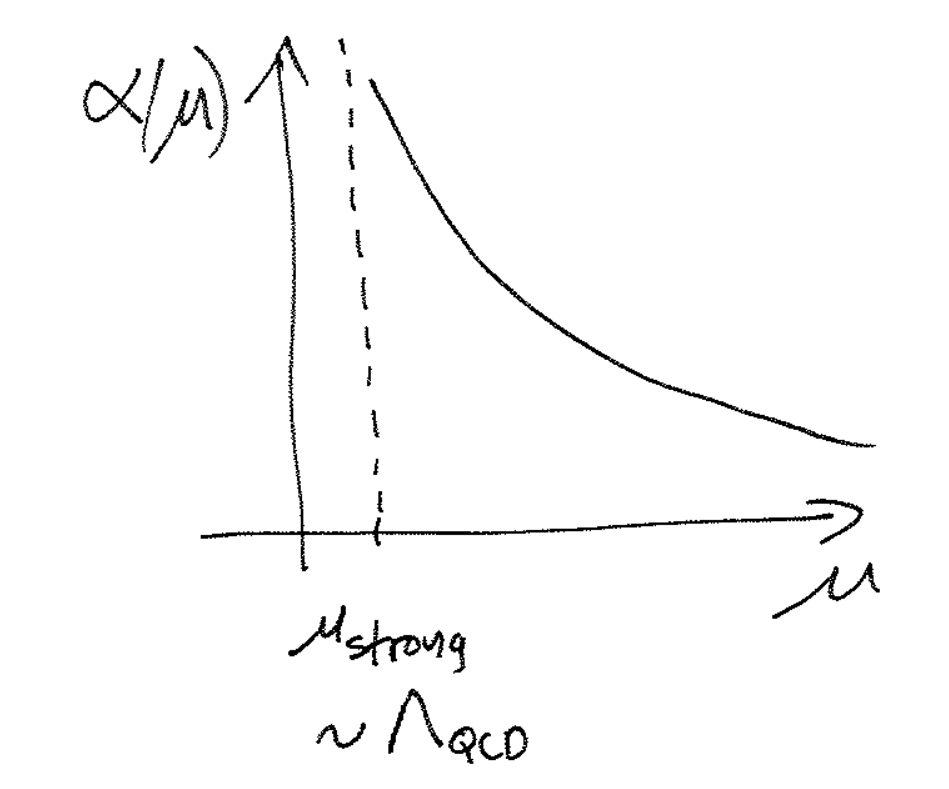}
\caption{Sketch of the  $beta$-function of QCD (left) and the
  corresponding running coupling constant $\alpha(\mu)$ (right).
\label{fig:run-qcd}}
\end{figure}

\section{Walking Technicolor: {\it A fairy tale (but a good one, and
    an interesting one)}}
Here is what Walking\,\cite{WTC} TC (WTC) is supposed to do: dynamically
replace $\LTC^2/M$ for $\LTC^3/M^2$ in the ETC quark mass formula, so
that $m_q\sim C \LTC^2/M$. Using the same assumptions as before this
gives $m_q\lesssim 10$~GeV. Therefore in walking technicolor one can
account for the masses of quarks and leptons without running afoul of
FCNC bounds. Except for the top quark. Being extraordinarily heavy it
has to be dealt with separately (this is true also of many an SM
extension). New interactions must be introduced, \eg, topcolor
assisted technicolor,\cite{Hill:1994hp} but because of time and space
limitations we do not consider them further here.

Before discussing how WTC works let us review a more familiar
example. The left diagram in Fig.~\ref{fig:run-qcd} is a sketch of the
beta function of QCD. It is a negative, monotonically decreasing
function of $\alpha$.  The sketch on the right hand of the figure
shows the running coupling, that is, the solution to the RGE:
$\mu\partial\alpha/\partial\mu=\beta(\alpha)$. It is best to think of
this as starting from some $\alpha(\mu_0)\ll1$ and integrating the RGE
towards the region $\mu<\mu_0$. In physics lingo, fix the value of
$\alpha$ to be perturbatively small in the UV and figure out what it
flows to in the IR. The sketch shows that $\alpha$ gets increasingly
large as it flows towards the IR and eventually at some scale
$\mu=\mu_{\text{strong}}\sim\Lambda_{\text{QCD}}$ becomes strong.

\begin{figure}
\includegraphics[height=6cm]{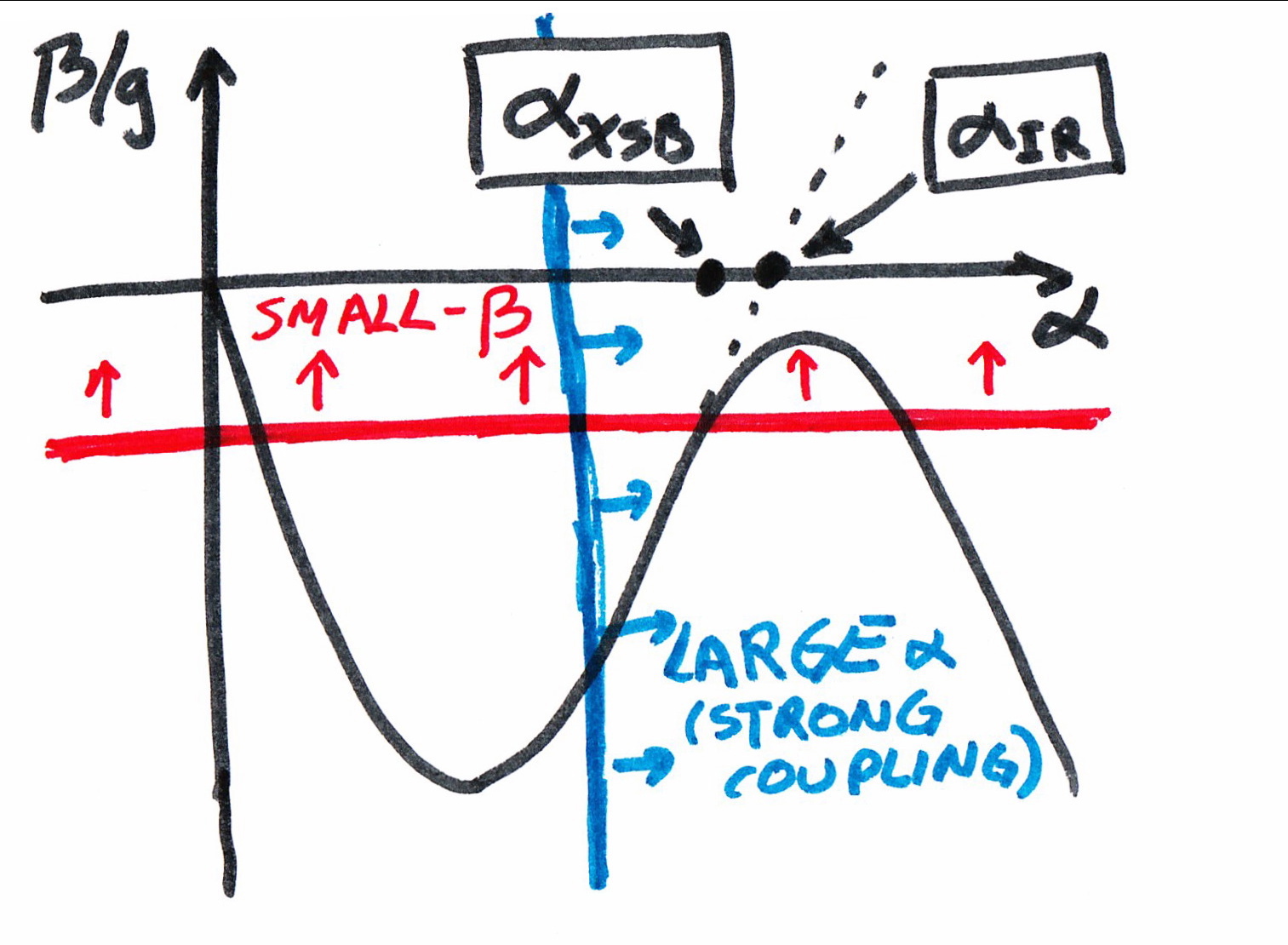}\hspace{-0.9cm}
\includegraphics[height=6cm]{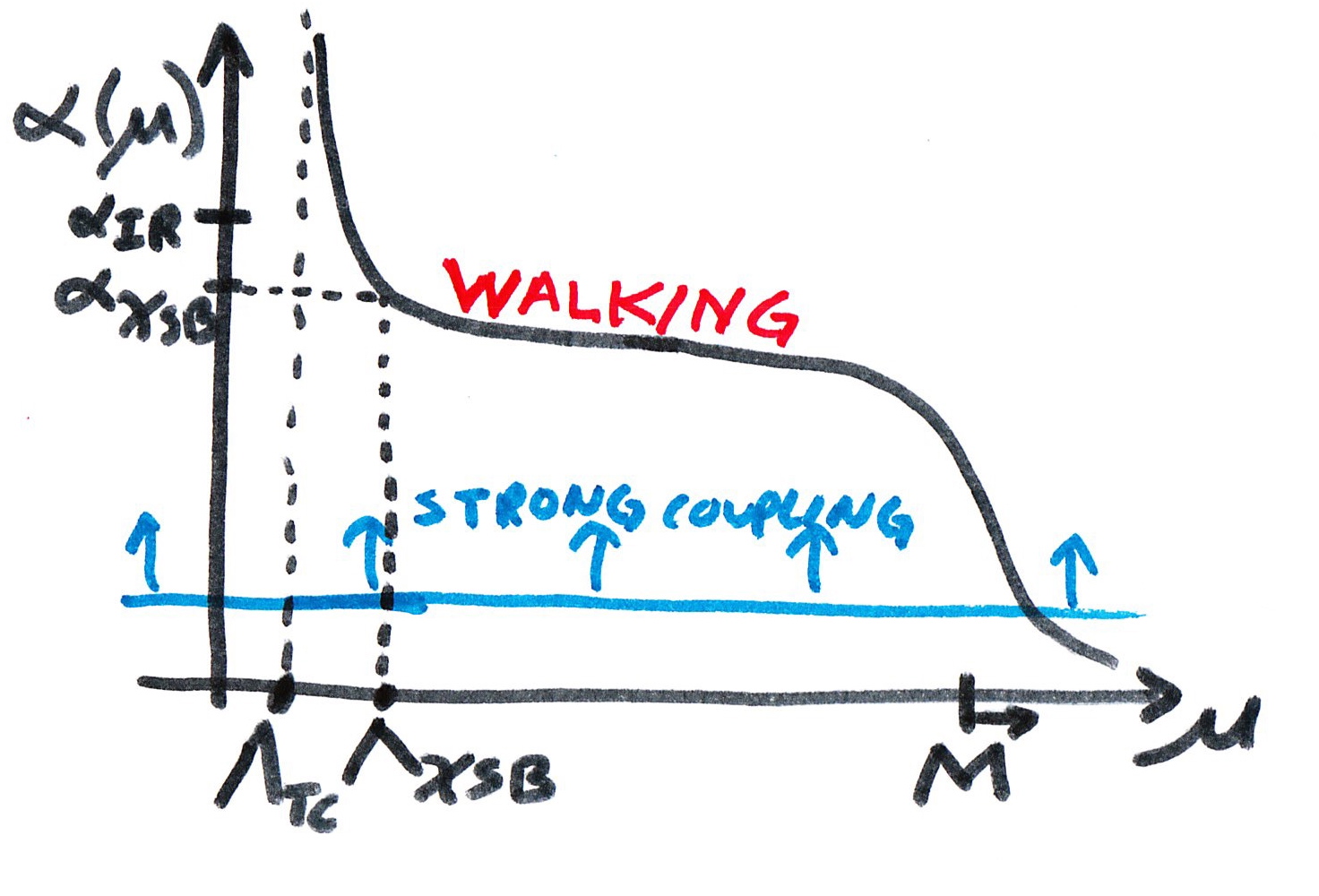}
\caption{Sketch of the beta function of WTC (left) and the
  corresponding running coupling constant(right). 
\label{fig:run-wtc}}
\end{figure}

Fig.~\ref{fig:run-wtc} shows a hypothetical $\beta$-function (or, should
I say, the $\beta$-function for a hypothetical theory?) and the
corresponding solution to the RGE. Starting from $\alpha(\mu_0)\ll1$
the movie script is as follows: $\alpha$ runs slowly because $\beta$
is small (close to the origin of the left sketch in
Fig.~\ref{fig:run-wtc}); eventually $\beta$  is large and $\alpha$
runs fast to large values; as $\alpha\sim1 $ increases further $\beta$
decreases to a small value, and the rate of change of $\alpha$ gets to
be puny. At this stage of the movie we say that $\alpha$ is
``walking'' rather than ``running.'' Now, in fact, it looks like this
would go on indefinitely: if we extrapolate the $\beta$-function it
appears as if it would cross the abscissa at the point labeled
``would-be IR fixed point, $\alpha_{\text{IR}}$'' in the figure. But
although $\alpha$ is barely changing, it  reaches and then
exceeds a critical coupling, $\alpha_{\chi\text{SB}}$ that triggers
chiral symmetry breaking ($\chi\text{SB}$). Note that by necessity
$\alpha_{\chi\text{SB}}<\alpha_{\text{IR}}$. As a result of
$\chi\text{SB}$   techniquarks acquire mass, which in turn produces the
sudden change in the $\beta$-function that, in the last frame of the
movie,  drives $\alpha$ arbitrarily
large, much like the  end state of QCD.  Finally, as
indicated in the sketch we assume the scale
$M$ of ETC is large, in the region before the coupling
walks.\footnote{
There are other ways of achieving this effect. All
  that is needed really is that $\beta$ be small when $\chi$SB occurs.\cite{WTC}} 

Now, here is how this helps. The effective ETC interaction of
\eqref{eq:ETC} is an RGE invariant, but both the coefficient $C$ and the
operator $\bar QQ$ are rapidly varying functions of $\mu$. The quark
mass term in the Lagrangian can be written as
\begin{equation}
\mathcal{L}_{\text{mass}}= \frac{C(\mu)}{M^2}\vev{\bar  QQ}(\mu)\bar qq
=\frac{C(M)}{M^2}\vev{\bar  QQ}(M)\bar qq
=\frac{C(\LTC)}{M^2}\vev{\bar QQ}(\LTC)\bar qq\,.
\end{equation}
The three forms of the equation are equivalent, by RGE invariance.
$C(M)$ is readily computed by ``matching'' to the full ETC, and does
not involve the scale $\LTC$ at all. On the other hand $\vev{\bar
  QQ}(\LTC)\sim\LTC^3$ from TC dynamics and does not involve the scale
$M$ at all. To compute the quark mass one needs either $C(\LTC)$ or $\vev{\bar
  QQ}(M)$ both of which involve implicit dependence on
$\LTC/M$. Fortunately, the RGE tells us what this dependence
is. The dominant effect is from the region where $\alpha$ is walking
and large,
producing a nearly constant and large anomalous dimension $\gamma_*$. Hence
\begin{equation}
C(\LTC)=\left(\frac{M}{\LTC}\right)^{\gamma_*}C(M)\,.
\end{equation}
Furthermore, it is argued that $\gamma_*\approx1$.\cite{Cohen:1988sq}
So we read off $m_q\sim C(M)\LTC^{3-\gamma_*}/M^{2-\gamma_*}\sim
C(M)\Lambda^2/M$ as advertised at the top of the section.

As a bonus, this mechanism raises the mass of technipions in
non-minimal TC. Recall the ETC interaction $C''/M^2(\bar Q Q)^2$
induces a technipion mass. In WTC this interaction will be enhanced,
much like the quark masses. In fact, since this involves the square of
$\bar Q Q$ we venture a guess that the anomalous dimension for $C''$ is
now closer to $2\gamma_*\approx2$. This means that $(\bar Q Q)^2$ is
effectively a dimension 4 operator, \ie, it is near marginal. If that
is the case there is no sense in which the symmetry was even
approximate and  the technipions are heavy, roughly as heavy as other
technimesons (like technirhos).  This plays a central role in next
section: $m_{\rho_T}<2m_{\pi_T}$ implies narrow, and hence more easily
detected, technirho.\cite{Lane:1989ej}

\paragraph{Speculation And Conjecture} WTC is speculation on the
nature of the universe based on a conjecture, namely, that a class of
quantum field theories behave as wanted. We don't have any examples of
such a theory. Yet, there is no no-go theorem. The conjecture  may be
affirmed or repudiated by studies of  QFT on the Lattice. The
speculation may be confirmed by the LHC. If that were the case the
future of particle physics may be much  like learning about strong
interactions from experiment, all over again. The new technistrong
interaction would necessarily behave very differently from QCD and
possibly be
less amenable than QCD to analytic studies.

\paragraph{EW Precision Tests} Since the EW sector in TC models is
strongly coupled there does not exist a reliable way to calculate
radiative corrections ($S$ and $T$ parameters). The exception is
minimal TC, since one can scale up from QCD measured
quantities. Minimal TC is ruled out by EW precision data.  So what? It is also
ruled out by quark masses/FCNCs, light technipions, etc.  To repeat
from the previous paragraph, the WTC theory is surely very unlike QCD. 
Low energy data offers no guidance as to how to calculate EW
corrections.

\section{WTC  at the  LHC:  the TCSM}
If WTC is discovered then there will be a long period of learning from
experiment about this new strongly coupled dynamics. The prerequisite
is the discovery, and we shall focus on it for now. In spite of
lacking detailed 
understanding of the dynamics, there are rough requirements on it
that may guide a search. This is the approach adopted by
the TC working group of the Les Houches Report.\cite{Brooijmans:2010tn}
I will summarize here their work, but in passing will butcher
it,\footnote{So I will not give you the detailed parameters used, what
  the analysis strategy is for each mode, nor will I talk about signal
  and event selection or  background estimation.} so readers
interested in details are advised to proceed directly to the source.

The idea is to focus on the least dynamics specific and most model
independent features of WTC. Since we must have $SU(2)_W$ we can use
it as the analogue of isospin to classify the spectrum: $\pi_T,
\rho_T, a_T, f_T,\pi'_T,\ldots$ We make the plausible assumption that the
large symmetry breaking effect of $(\bar Q Q)^2$ makes $\pi_T$
sufficiently heavy that the other low mass mesons cannot decay into
technipions, $m_{\rho_T}<2m_{\pi_T}$,  $m_{\omega_T}<3m_{\pi_T}$,
etc. Hence there are many {\it narrow} states. This set-up is called
``Low Scale TC'' or, alternatively, the ``TC Straw Man'' (TCSM). 

The Les Houches report presents studies of Drell-Yan production of
technimesons in $pp$ collisions at 10~TeV energy searched for in
$\rho_T^\pm\to W^\pm Z^0$, $(\rho_T,a_T)\to W\gamma$, $\omega_T\to\gamma
Z^0\to\gamma\ell^+\ell^-$  and
$(\rho_T^0,a_T^0,\omega_T)\to\ell^+\ell^-$. The last process would be
forbidden if the electric charges of the up and down techniquarks
coherently vanish, $Q_U+Q_D=0$; for that study it is assumed that
$Q_U+Q_D=1$. If $Q_U+Q_D=0$ the discovery channel for $\omega_T$ may
well be $\omega_T\to\gamma Z^0$. The cases studied (assumed mass
spectra) and gross results are presented in Table
\ref{tabl}. $\text{Br}(W/Z\to (e,\mu)X)$ are included.  $\sigma(e^+
e^-)$ includes SM plus signal production integrated over approximately
$M_{\rho_T,\omega_T} - 25$~GeV to $M_{a_T} + 25$~GeV; for orientation,
the cross section in the SM alone is in parentheses. As you can see,
many of these events will be produced in 1~$\text{fb}^{-1}$ of data.

\def\tro{\rho_T}\def\tom{\omega_T}
\def\tpi{\pi_T}\def\ta{a_T}

%
%
\begin{table}[!ht]
\begin{center}{
  \begin{tabular}{|c|c|c|c|c|c|c|c|c|c|}
  \hline
 Case & $M_{\tro,\tom}$ & $M_{a_T}$ & $M_{\tpi}$ && $\sigma(W^\pm Z^0)$ & $\sigma(\gamma
 W^\pm)$ & $\sigma(\gamma Z^0)$ & $\sigma(e^+e^-)$ \\
  \hline\hline
 1a & 225 & 250 & 150 && 230 & 330 & 60 & 1655 (980)\\
 1b & 225 & 250 & 140 && 205 & 285 & 45 & 1485 (980)\\
 2a & 300 & 330 & 200 &&  75 & 105 & 11 &  425 (290)\\
 2b & 300 & 330 & 180 &&  45 &  85 &  7 &  380 (290)\\
 3a & 400 & 440 & 275 &&  22 &  40 &  4 &  130  (90)\\
 3b & 400 & 440 & 250 &&  14 &  35 &  3 &  120  (90)\\
   \hline\hline
 \end{tabular}}
\caption{Technihadron masses (in GeV) and 
    approximate signal cross sections  (in fb) for $pp$ collisions at
  $\sqrt{s} = 10$~TeV taken from the 2009 Les Houches report (where
  details of  parameters  of the TCSM can be
  found). 
   \label{tabl}}
 \end{center}
 \end{table}
%
%

\section{Other Stuff and Final Remarks}
The title I had been assigned when invited to give this talk was
``Strong Dynamics and New Electroweak Models.'' As I explained, to
avoid overlap with other, earlier talks I decided to concentrate on
``Strong EW Symmetry Breaking.'' I would like to close with a few
words about other possibilities (that do not overlap previous
talks!). 

In WTC the bilinear $\bar Q_LQ_R$ plays the role of $H$, the SM's
higgs doublet. As we saw, it's scaling dimension is 2 and therefore
the quark/lepton mass terms are irrelevant operators (dimension
5). {\it Conformal TC}\,\cite{Luty:2004ye} proposes that $H\sim \bar
Q_LQ_R$ has dimension close to 1 while $H^2\sim (\bar Q_LQ_R)^2$ has
dimension close to 4. If this were the case both the quark/lepton mass
terms (\eg, $H\bar q d$) and the ``higgs mass'' $H^2$ would be
marginal operators. It would follow that their dependence on cutoff is
logarithmic and again there would be no hierarchy or fine tuning
problem. Moreover, accommodating a top quark mass should be as simple
as in the standard model. Unfortunately there is some evidence against
this arrangement of anomalous dimensions.\cite{Rattazzi:2008pe}

5D models give a means to study the idea that the masses of $W$ and
$Z$ arise from internal structure. That is, if you believe duality
applies here. Take for example a model with the $W/Z$ localized on the
IR brane, and let boundary conditions break gauge symmetry
explicitly.\cite{Cui:2009dv} The 5D/4D dictionary then gives a
higgsless 4D dual with composite $W/Z$. Just like techniresonances in
TC this model has 4D resonances, the duals of the Kaluza-Klein 5D
modes. Even in this setup, the top mass remains a problem.

Strong EW symmetry breaking is very appealing. I believe news of its
demise are premature, driven mostly by our frustration with our
inability to calculate. What we do know is that it is not a scaled up
version of QCD. These comments apply equally to all alternatives to 
TC. So the strategy for progress is to (i) search and discover, (ii)
study in detail and (iii) build a model and learn about strong
dynamics. In that order!

The TCSM is a good example of how to engage this strategy. It is not a
model but a template for searching for WTC, making a number of
assumptions (hopefully minimal and reasonable). Similar efforts that
can cover alternative strong EW breaking scenarios are needed.

Can strong EW symmetry breaking be excluded? Or, at what point after
not finding evidence do we give up on the idea? Lack of a definite
answer by practitioners has often frustrated others, prompting some to
facetiously ask\,\cite{Ellis:1980hz} ``Can One Tell Technicolor from a
Hole in the Ground?'' I think the question is no different than for
other models (when do we give up on SUSY?). This is not a productive
avenue of thought, particularly if you believe that many discoveries
await us in the dawn of LHC physics. As of this writing the CMS and
ATLAS experimental collaborations posted integrated luminosities of
about $700~\text{nb}^{-1}$ each. That's 1400 down, 998,600 to
go. Looking forward to it!

\section*{Acknowledgments}
Thanks to Julius Kuti and Ken Lane for helpful
discussions and CERN-TH for their hospitality. Work
supported in part by U.S. DoE under Contract
No. DE-FG03-97ER40546

\end{document}